\documentclass[aps,prl,twocolumn,superscriptaddress]{revtex4}%
\usepackage{epsfig,dsfont,amssymb,amsmath,amsthm,amsfonts,amsbsy,mathrsfs}
\usepackage{graphicx}
\usepackage{amsmath}
\usepackage{amssymb}%
\usepackage[utf8x]{inputenc}
\usepackage{graphicx}
\usepackage{dcolumn}
\usepackage{bm}
\usepackage{xcolor}
\usepackage{amsmath}
\usepackage{chngcntr}
\usepackage{nicefrac}
\usepackage[colorlinks,linkcolor=blue,hyperindex,CJKbookmarks]{hyperref}
\begin{document}

\title{Prevailing charge order in overdoped cuprates beyond the superconducting dome}

\author{Qizhi Li}\thanks{These authors contributed equally to this work.}
\affiliation{International Center for Quantum Materials, School of Physics, Peking University, Beijing 100871, China}

\author{Hsiao-Yu Huang}\thanks{These authors contributed equally to this work.}
\affiliation{National Synchrotron Radiation Research Center, Hsinchu 30076, Taiwan}

\author{Tianshuang Ren}\thanks{These authors contributed equally to this work.}
\affiliation{Interdisciplinary Center for Quantum Information, State Key Laboratory of Modern Optical Instrumentation, and Zhejiang Province Key Laboratory of Quantum Technology and Device, Department of Physics, Zhejiang University, Hangzhou 310027, China}

\author{Eugen Weschke}
\affiliation{Helmholtz-Zentrum Berlin für Materialien und Energie, Berlin, Germany}

\author{Lele Ju}
\affiliation{Interdisciplinary Center for Quantum Information, State Key Laboratory of Modern Optical Instrumentation, and Zhejiang Province Key Laboratory of Quantum Technology and Device, Department of Physics, Zhejiang University, Hangzhou 310027, China}

\author{Changwei Zou}
\affiliation{International Center for Quantum Materials, School of Physics, Peking University, Beijing 100871, China}

\author{Shilong Zhang}
\affiliation{International Center for Quantum Materials, School of Physics, Peking University, Beijing 100871, China}

\author{Qingzheng Qiu}
\affiliation{International Center for Quantum Materials, School of Physics, Peking University, Beijing 100871, China}

\author{Jiarui Liu}
\affiliation{Department of Physics and Astronomy, Clemson University, Clemson, South Carolina 29631, USA}

\author{Shuhan Ding}
\affiliation{Department of Physics and Astronomy, Clemson University, Clemson, South Carolina 29631, USA}

\author{Amol Singh}
\affiliation{National Synchrotron Radiation Research Center, Hsinchu 30076, Taiwan}

\author{Oleksandr Prokhnenko}
\affiliation{Helmholtz-Zentrum Berlin für Materialien und Energie, Berlin, Germany}

\author{Di-Jing Huang}
\affiliation{National Synchrotron Radiation Research Center, Hsinchu 30076, Taiwan}

\author{Ilya Esterlis}
\affiliation{Department of Physics, Harvard University, Cambridge, Massachusetts 02138, USA}

\author{Yao Wang}
\affiliation{Department of Physics and Astronomy, Clemson University, Clemson, South Carolina 29631, USA}

\author{Yanwu Xie}
\affiliation{Interdisciplinary Center for Quantum Information, State Key Laboratory of Modern Optical Instrumentation, and Zhejiang Province Key Laboratory of Quantum Technology and Device, Department of Physics, Zhejiang University, Hangzhou 310027, China}

\author{Yingying Peng}
\email{yingying.peng@pku.edu.cn}
\affiliation{International Center for Quantum Materials, School of Physics, Peking University, Beijing 100871, China}

\date{\today}

\begin{abstract}
The extremely overdoped cuprates are generally considered to be Fermi liquid metals without exotic orders, whereas the underdoped cuprates harbor intertwined states. Contrary to this conventional wisdom, using Cu $L_3$ edge and O $K$ edge resonant x-ray scattering, we reveal a charge order (CO) in overdoped La$_{2-x}$Sr$_x$CuO$_4$ (0.35 $\leq$ x $\leq$ 0.6) beyond the superconducting dome. This CO has a periodicity of $\sim$ 6 lattice units with correlation lengths of $\sim 3 - 20$ lattice units. 
It shows similar in-plane momentum and polarization dependence and dispersive excitations as the CO of underdoped cuprates, but its maximum intensity differs along the c-direction and persists up to 300 K. This CO cannot be explained by either the Fermi surface instability or the doped Hubbard model and its origin remains to be understood. Our results suggest that CO is prevailing in the overdoped metallic regime and superconductivity emerges out of the CO phase upon decreasing hole carriers.

\end{abstract}

\maketitle

High-temperature superconductivity (SC) is a great surprise in quantum materials and its mechanism remains a puzzle. In the past 36 years, studies on cuprate superconductors primarily have been focused on the underdoped and optimally doped regions — close to the Mott-insulating state of the phase diagram \cite{wenxiaogang,keimer}. It has been a longstanding challenge to understand how the versatile phenomena exhibited in these materials, such as the pseudogap (PG) and strange metal states, together with a plethora of exotic electronic orders, coexist and compete with superconductivity \cite{keimer}. On the contrary, the overdoped region is generally considered to be a conventional Fermi liquid (FL) and less affected by the doped-Mott-insulator scenario \cite{wenxiaogang}. However, the doping and temperature dependencies of the superfluid density are incompatible with the standard BCS description, suggesting phase fluctuations in the overdoped region \cite{bovzovic_LSCO_superfluiddensity}. These fluctuations are characterized by photoemission as preformed Cooper pairs well above T$_c$ \cite{ShenZXODBi2212} and magnetic fluctuations persistent in the extremely overdoped regime \cite{paramagnon}. Moreover, ferromagnetism has been discovered in overdoped (Bi,Pb)$_2$Sr$_2$CuO$_{6+\delta}$ (Bi2201) \cite{FM_Bi2201} and beyond the superconducting dome in La$_{2-x}$Sr$_x$CuO$_4$ \cite{FM_overdopedLSCO}. 
These discoveries suggested that strong electronic correlations remain present in the overdoped cuprates.

Charge orders are ubiquitous in underdoped cuprates \cite{tranquada1995evidence,abbamonte2005spatially,Ghiringhelli2012,chang2012direct,cominCO,YYPBi2201,caipeng,LSCOSDWCDW,0.21CDW,keimer} and has enigmatic interactions with SC \cite{Ghiringhelli2012,chang2012direct} and PG \cite{cominCO}; however, there is still no consensus on the underlying mechanism with on-going debates including the real-space electronic correlation scenario \cite{caipeng,0.21CDW} versus the momentum-space instability scenario \cite{cominCO}. The latter has been challenged by one resonant X-ray scattering study that revealed charge order in heavily overdoped Bi2201 whose Fermi surface lacks the nesting features \cite{YYPBioverdopedCDW}. The CO provides a route by which to test how close the overdoped systems are to a FL, and also quantify the underlying fluctuations. Therefore, it is of great importance to assess the universality of charge ordering in overdoped cuprates, which will not only have the potential to elucidate the mechanism of CO instabilities but also provide a valuable perspective to understand the complex phase diagram coming from the overdoped regime.

\begin{figure*}
\centering\includegraphics[width = 0.9\textwidth]{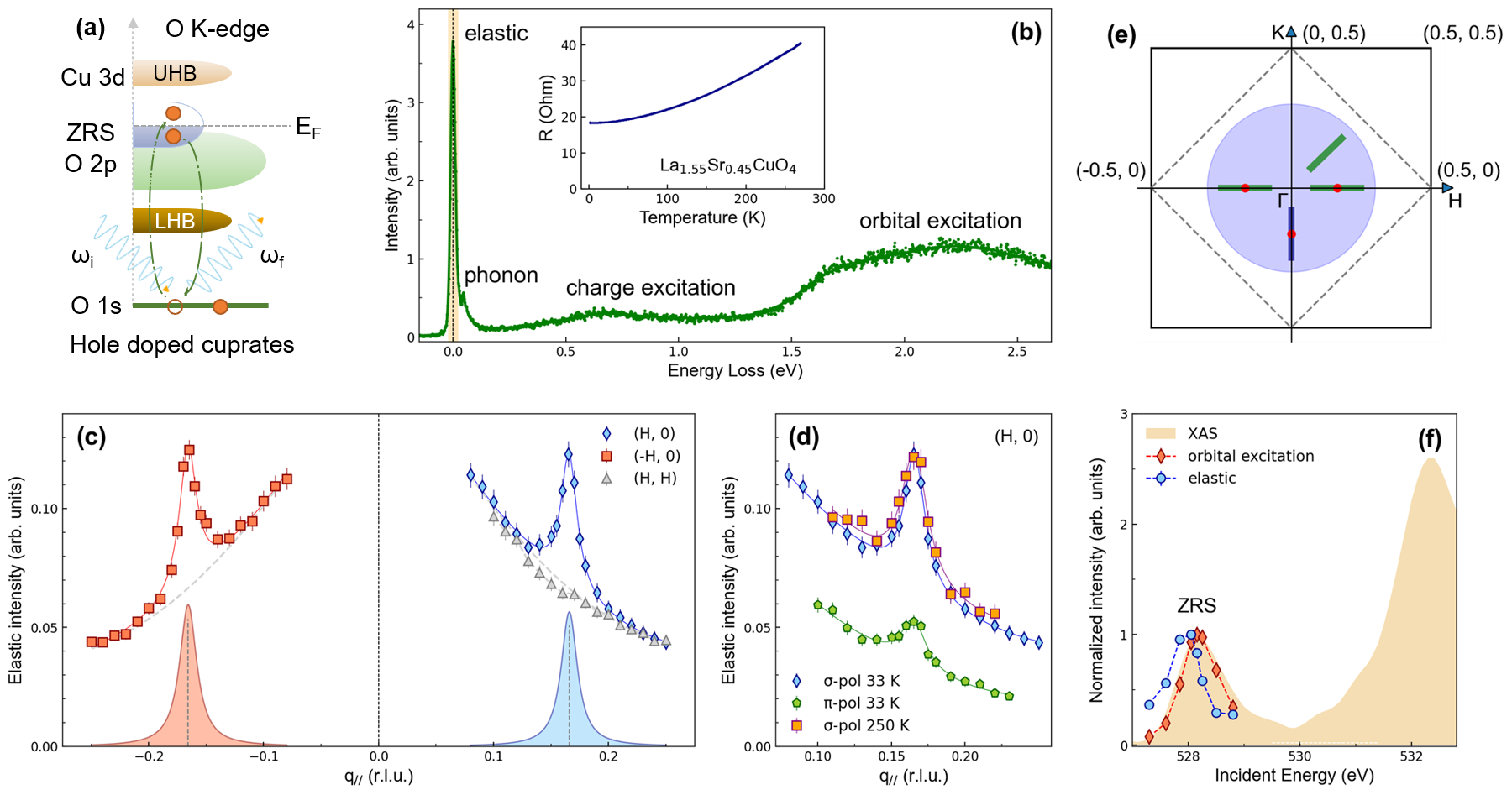}
\caption{\label{fig:COLSCO(0.45)} Observation of charge order by RIXS in overdoped metallic La$_{2-x}$Sr$_x$CuO$_4$ ($x$ = 0.45). (a) Schematic plot of RIXS process at O \emph{K}-edge. (b) A typical RIXS spectrum at $\bf q_{||}$ = (0.12, 0), displaying the elastic peak, phonon, charge and orbital excitations. Inset: resistance curve displaying the metallic nature of the sample. (c) Integrated intensity of elastic peaks for positive and negative (H, 0), and (H, H) directions, using $\sigma$ polarization. Red and blue curves are Lorentzian peak fits to the data with a polynomial background (gray dashed lines). (d) Polarization measurements with $\sigma$- and $\pi$-polarized light, collected at 33 K and 250 K. (e) Reciprocal-space image. The Blue shaded region is the accessible momentum-transfer range of O \emph{K}-edge. The green and blue lines indicate the momentum cuts, and red dots indicate the observed CO peaks. (f) XAS spectra near the Zhang-Rice Singlet (ZRS) absorption peak with $\sigma$-polariztion at normal incidence. Incident energy dependence of the integrated intensity of elastic peak and orbital excitation, normalized to the value at the ZRS peak. }
\end{figure*}

In this Letter, we report the existence of charge order in heavily overdoped La$_{1.55}$Sr$_{0.45}$CuO$_{4}$ thin films beyond the superconducting dome (sample growth and characterizations are described in \emph{Supplementary} Sec.1). The energy resolution of oxygen \emph{K}-edge resonant inelastic x-ray scattering (RIXS) is $\sim$ 30 meV at 41A Taiwan photon source. Figure~\ref{fig:COLSCO(0.45)}(a) displays the schematic electronic structure of hole doped cuprates and illustrates the resonant absorption as well as the scattering processes at oxygen \emph{K}-edge (1\emph{s} $\rightarrow$ 2\emph{p}).
The Zhang-Rice singlet (ZRS), originating from the hybridization between oxygen ligands and Cu $3d_{x^2-y^2}$ orbitals, manifests as a pre-edge peak in the x-ray absorption spectra (XAS) \cite{CTchenXAS}.
Figure~\ref{fig:COLSCO(0.45)}(b) exemplifies the observed excitations in a metallic LSCO with $x$ = 0.45 (resistivity shown in the inset of Fig.~\ref{fig:COLSCO(0.45)}(b)), including the elastic peak, phonons ($\sim0.05$ eV), charge excitations ($\sim0.6$ eV) \cite{plasmonZhouKJ}, and oxygen orbital excitations ($>1.5$ eV). Given the relevance to CO \cite{Ghiringhelli2012,dispersiveCDW,JieminCDW}, we focus on the elastic peak and low-energy phonons below.

The elastic scattering displays a prominent peak at the planar wavevector $q_{||}$ = (0.165, 0).
The full width at half-maximum (FWHM) of the peak is $\sim$ $0.017 \pm 0.002$ $r.l.u.$ with a correlation length of $\sim 70.8~\AA$. This feature is symmetric along both (H, 0) and (-H, 0) directions but absent along diagonal (H, H) directions (Fig.~\ref{fig:COLSCO(0.45)}(c)) (see RIXS map in \emph{Supplementary} Fig. S5). 
Its intensity is more pronounced for $\sigma$-polarization than for $\pi$-polarization (Fig.~\ref{fig:COLSCO(0.45)}(d)), suggesting a predominant $d_{x^2-y^2}$ character of dopants forming the charge order. And this peak is nearly unchanged at temperatures as high as 250 K (Fig.~\ref{fig:COLSCO(0.45)}(d)), which is similar to the CO in overdoped Bi2201 \cite{YYPBioverdopedCDW}. Moreover, this temperature-independent behavior is in line with the short-range high-temperature charge fluctuations (CDF) in the underdoped regions \cite{GiacomoScienceCDF}.
We also checked the resonant behaviour of this peak to reveal whether it originates from a modulation of the valence electrons. Figure~\ref{fig:COLSCO(0.45)}(f) shows the intensity of the integrated elastic peak and orbital excitation as a function of incident photon energy compared with XAS. The orbital excitation follows the XAS spectrum, while the integrated elastic peak exhibits a clear resonance with a maximum slightly below the ZRS pre-peak. That is because XAS and scattering signals correspond to the imaginary and real parts of the atomic form factors, respectively. This trend is closely analogous to the behaviour of COs in underdoped cuprates \cite{abbamonte2005spatially,JieminCDW}.
We have further excluded this peak as a trivial superstructure from hard x-ray scattering measurements (see \emph{Supplementary} Fig. S4). 
Thus, we will refer to this feature as a CO peak below.
Notably, here the CO wavevector $q_{\rm CO}$ = (0.165, 0) is much smaller than that at underdoped LSCO samples — typically $q_{\rm CO}\sim$ (0.23, 0) \cite{0.21CDW,EPCRIXS,npjLSCOCDW}. It nevertheless is close to the value in overdoped Bi2201 with $q_{\rm CO}\sim$ (0.13, 0) \cite{YYPBioverdopedCDW}, suggesting that charge instability with relatively long wavelength may prevail in heavily overdoped cuprates. 

\begin{figure}[t]
\centering
\includegraphics[width = 0.9\columnwidth]{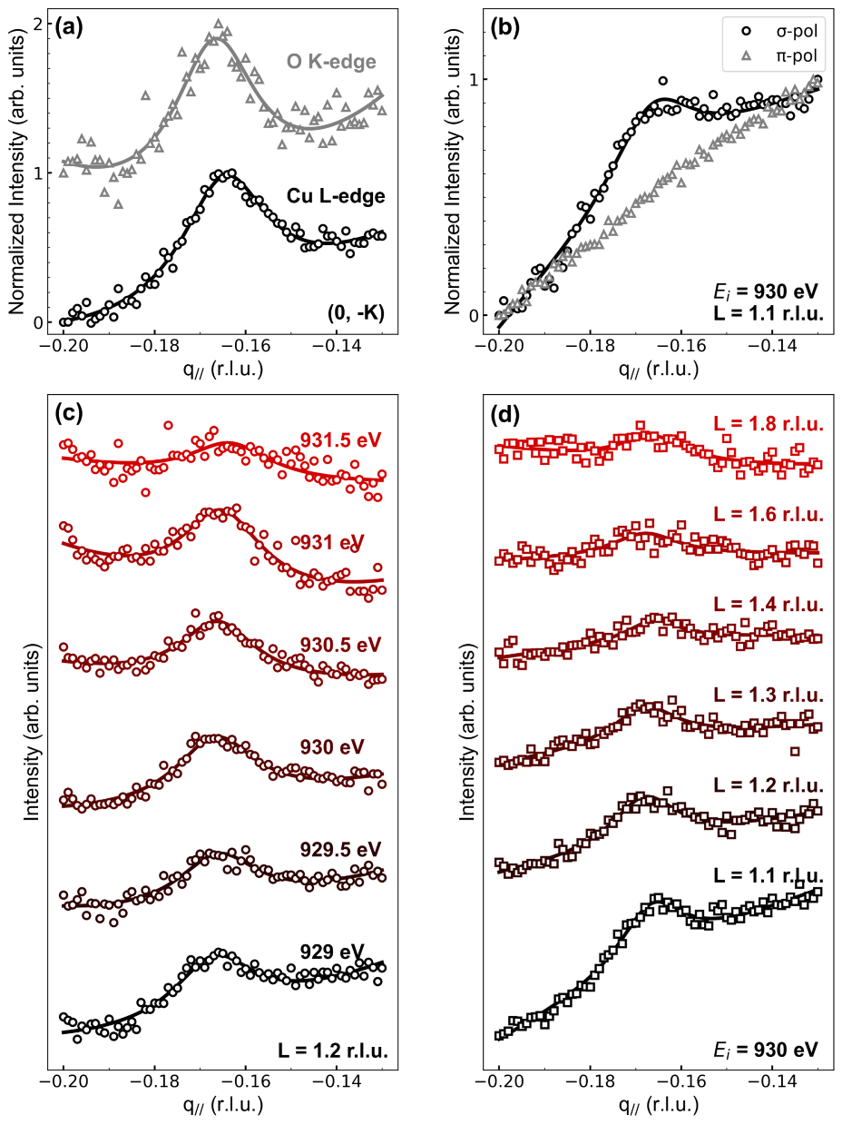}
\caption{\label{fig:REIXS} Cu $L_3$-edge and O \emph{K}-edge REIXS studies of charge order in La$_{1.55}$Sr$_{0.45}$CuO$_4$. (a) Observation of charge order along (0, -K) direction at both Cu $L_3$-edge and O \emph{K}-edge, offset is applied for clarity. (b) Polarization dependence of the charge order peak, collected at 930 eV and \emph{L} = 1.1 $r.l.u.$. (c) Detuning measurements near the Cu $L_3$-edge after self-absorption correction (see \emph{Supplementary}). (d) \emph{L}-dependence of charge order within the accessible range of [1.1, 1.8] $r.l.u.$ at 930 eV, collected with $\sigma$-polarization.}
\end{figure}

To investigate whether this CO signal originates from CuO$_2$ planes, we also perform Cu \emph{L}-edge resonant energy integrated x-ray scattering (REIXS) measurements at Helmholtz-Zentrum Berlin. As shown in Fig.~\ref{fig:REIXS}(a), we have measured the CO peak along (0, -K) direction (blue cut in Fig.~\ref{fig:COLSCO(0.45)}(e)) at both Cu \emph{L}-edge and O \emph{K}-edge and found it identical at the two edges, suggesting the strong hybridization between the Cu-3$d$ and the O-2$p$ orbitals. This result also demonstrates that the CO is bidirectional along $H$- and \emph{K}-directions. 
Moreover, the CO peak measured by REIXS overlaps very well with the energy-integrated RIXS result (see \emph{Supplementary} Fig. S7), proving the consistency of the two experimental techniques. We again observe that the CO peak is stronger at $\sigma$-polarization than $\pi$-polarization at Cu \emph{L}-edge (Fig.~\ref{fig:REIXS}(b)) in favor of a charge origin \cite{Ghiringhelli2012}. 
Figure~\ref{fig:REIXS}(c) shows the incident energy dependence of CO peak across Cu $L_3$-edge, which does not display a clear resonant behavior due to a relatively poor energy resolution of $\sim$ 1.3 eV of REIXS measurement. By selecting the incident x-ray energy at 930 eV with a prominent CO peak, we further investigate the \emph{L} dependence of CO (Fig.~\ref{fig:REIXS}(d)). We observe that the CO peak maximizes at \emph{L} = 1.1 $r.l.u.$ with smaller \emph{L} value inaccessible, which is close to an integer \emph{L} value. This is in sharp contrast to the behavior of CO in underdoped LSCO \cite{npjLSCOCDW}, which has a maximum at half-integer \emph{L} due to the modulation of stripes along the c direction \cite{CDW2c}. The different \emph{L} behaviour may relate to the disappearance of spin glass behavior beyond the critical doping of the pseudogap phase ($x_c \sim 0.19$) in LSCO \cite {frachet2020hidden}, which is favoured by charge-stripe ordering. It is worth noting that CO in underdoped YBa$_2$Cu$_3$O$_{6.67}$ can be enhanced by suppressing SC under magnetic field \cite{YBCO3DCDW} or optical pump \cite{photoinducedCDW}, which also shows a maximum at integer \emph{L} values. The similar \emph{L}-dependent behaviors of COs induced by the high magnetic field, optical pump and metallic regime suggest that they may have the same origin and be a common feature in the normal state. 

\begin{figure}
\centering
\includegraphics[width = 0.95\columnwidth]{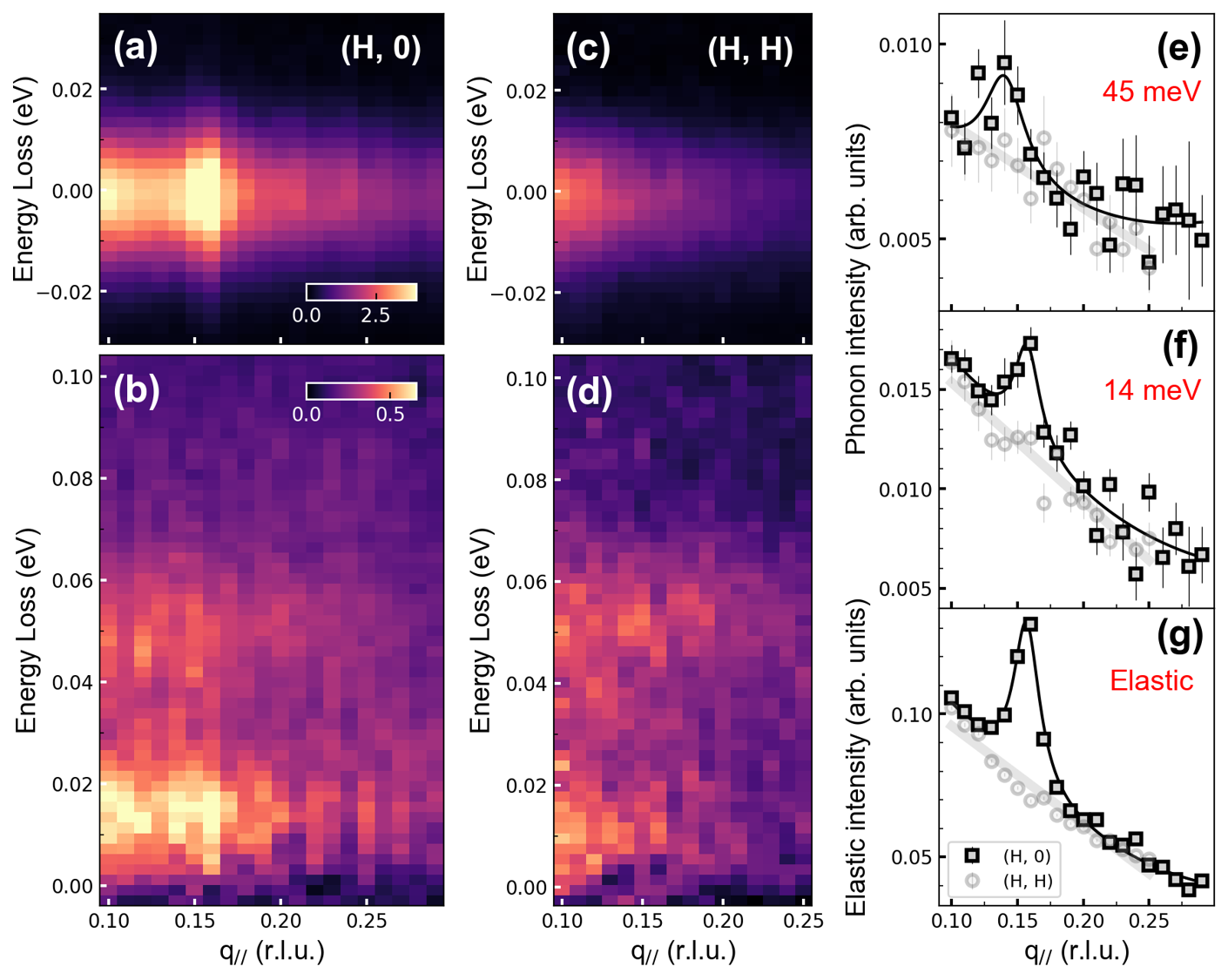}
\caption{\label{fig:phononLSCO(0.45)} Anisotropic momentum dependence of phonon intensity and dispersive CO excitations in La$_{2-x}$Sr$_x$CuO$_{4}$ ($x$ = 0.45). (a, c) The elastic intensity map and (b, d) the inelastic RIXS intensity map for visualizing the CO and phonon branches along (H, 0) direction and (H, H) direction, respectively. (e, f, g) Integrated intensity for buckling phonon, acoustic phonon and CO, respectively, along both directions. Details of the fitting are presented in Supplementary Materials. }
\end{figure}

To unveil the origin of CO it is crucial to investigate its collective excitations and the interplay between CO and phonons. In underdoped cuprates, it is widely observed that CO coexists with phonon intensity anomaly near the characteristic wavevector, accompanied by different magnitudes of phonon energy softening \cite{dispersiveCDW,JieminCDW,0.21CDW,melting,huangCDF,EPCRIXS}. These have been explained either by enhanced electron-phonon (e-ph) coupling \cite{EPCRIXS} or interference between collective charge fluctuations and phonons \cite{dispersiveCDW,JieminCDW,melting}. Exploiting the high energy-resolution of O \emph{K}-edge RIXS, we can probe CO and low-energy excitations at the same time. The RIXS map with better statistics and resolution ($\sim$ 25 meV) are visualized in Fig.~\ref{fig:phononLSCO(0.45)}(a). Notably, two phonon branches display pronounced enhancements of intensity near $q_{\rm CO}$, while the phonon energy softening is negligible here. To quantify the $q_{||}$ dependence of the phonons, we fit the inelastic part of RIXS spectra and reveal three features at $\sim$ 14 meV, $\sim$ 45 meV and $\sim$ 75 meV (See \emph{Supplementary} Fig. S6). They can be assigned to the acoustic, bond-buckling and bond-stretching phonon modes, respectively, in accord with a recent RIXS study on the optimally doped LSCO \cite{huangCDF}. 
Our intensity distribution curves show that acoustic (Fig.~\ref{fig:phononLSCO(0.45)}(e)) and buckling (Fig.~\ref{fig:phononLSCO(0.45)}(f)) branches are reinforced near $q_{\rm CO}$ (Fig.~\ref{fig:phononLSCO(0.45)}(g)), in concert with a proposed picture that dynamical CO excitations interfere with multiple phonon branches \cite{dispersiveCDW,JieminCDW,melting}. In stark contrast, no phonon intensity anomalies have been observed along (H, H) direction in the absence of CO (Fig.~\ref{fig:phononLSCO(0.45)}(c,d)), which suggests an intimate correlation between phonon and CO. We can further estimate the characteristic velocity of the dispersive CO excitations by connecting the maximal intensity anomaly of buckling mode at $q_A$ = 0.14 $r.l.u.$ with $q_{\rm CO}$ = 0.16 $r.l.u.$, giving a velocity of $\sim$ $1.4 \pm 0.4$ $eV \AA$. This matches very well with the CO velocity of $1.3 \pm 0.3$ $eV \AA$ in underdoped Bi2201 \cite{JieminCDW}. Our results indicate that dispersive CO excitations also exist in extremely overdoped regions, in analogy to those in underdoped regions.

\begin{figure}[t]
\centering\includegraphics[width = 0.95\columnwidth]{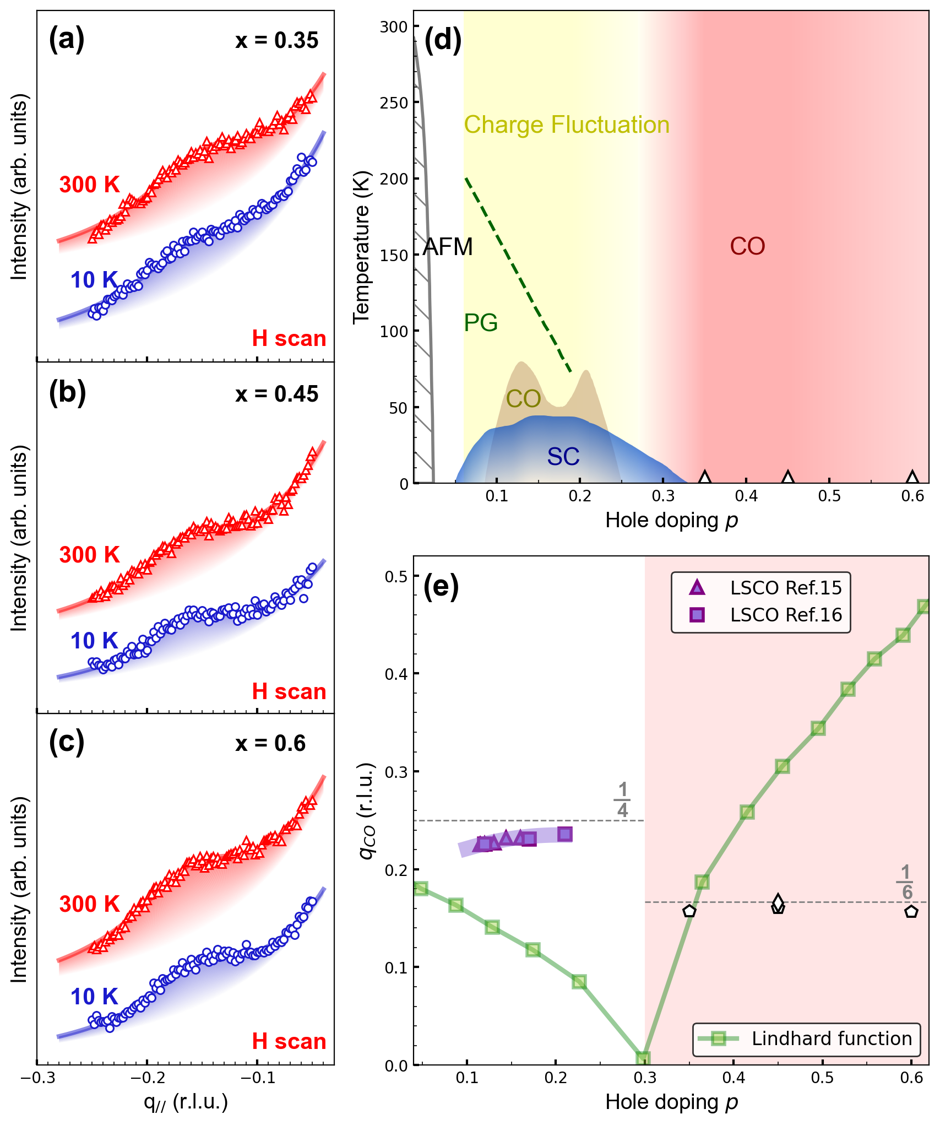}
\caption{\label{fig:phasediagram} Doping dependence of charge order in overdoped La$_{2-x}$Sr$_x$CuO$_{4}$ and the extended phase diagram. (a, b, c) CO peak profiles measured by Cu \emph{L}-edge REIXS in LSCO with $x$ = 0.35, 0.45, 0.6, respectively. The offset is applied for clarity. \emph{L} is fixed at 1.1 $r.l.u.$. The peak is nearly temperature independent up to 300 K. (d) The extended CO phase diagram of cuprates. It shows superconducting dome defined by T$_c$, antiferromagnetism (AFM) defined by T$_N$ \cite{PhysRevB_LSCOspin}, pseudogap (PG) determined from the Nernst coefficient \cite{LSCONernst2}, underdoped charge order and charge fluctuation \cite{npjLSCOCDW,0.21CDW,GiacomoScienceCDF} and overdoped charge order. (e) The doping dependence of the CO wave vector in LSCO and nesting vector obtained from Lindhard function. }
\end{figure}

To investigate the generality of this CO, we study the doping dependence of the CO on another batch of overdoped LSCO thin films ($x$ = 0.35, 0.45, 0.6) (Fig.~\ref{fig:phasediagram}). These films have a smaller thickness of 30 unit cells than the above-studied LSCO ($x$ = 0.45) with 50 unit cells. We observe that the CO appears at $x$ = 0.35 and gets more pronounced at $x$ = 0.45 and 0.6, which all persist from 10 K up to 300 K, as shown in Fig.~\ref{fig:phasediagram}, (a) to (c). We fit the REIXS spectrum with a Lorentz function for the peak and a Lorentz background from the specular tail, and the extracted peaks are shown in shaded areas. With smaller thickness, we observe that the CO peak displays a similar wavevector at $\sim0.166~r.l.u.$ but its correlation length becomes much shorter ($\sim$ 3 lattice units). This excludes the origin of CO driven from the substrate, while the thickness dependence remains to be understood. The observation of CO in extremely overdoped cuprates provides fresh insight for understanding the phase diagram (Fig.~\ref{fig:phasediagram}(d)). The presence of CO and AFM phases on the two sides of the superconducting dome is enlightening, suggesting that the unconventional superconductivity of cuprates can be regarded as an emergent phase out of either AFM or CO. This is consistent with the coexistence of short-ranged spin fluctuations and CO in the superconducting regime. Previous x-ray scattering studies report that the CO maximized at half-integer \emph{L} values would disappear at $x \sim 0.25$ in LSCO \cite{0.21CDW,npjLSCOCDW}. Here, the re-entrant CO exhibits distinct temperature and \emph{L} dependencies, as well as different in-plane wave vectors, suggesting different interactions in play compared to CO in the underdoped region.

We then discuss the possible origins of this re-entrant overdoped CO. First, its emergence at the extremely overdoped region implies that it does not correlate with the pseudogap phase that ends at $x_c \sim 0.19$ \cite{LSCONernst2}. We can also exclude the impact of the van Hove singularity, which was argued to cause the CDW phase in Bi2201\,\cite{YYPBioverdopedCDW}, since the Lifshitz transition occurs at much lower doping for LSCO ($x\sim0.2$) \,\cite{npjLSCOCDW}. 
Figure~\ref{fig:phasediagram}(e) shows the doping dependence of $q_{\rm CO}$ in LSCO \cite{LSCOSDWCDW,0.21CDW}. In the underdoped region, the CO wavevector increases with doping but close to $0.25~r.l.u.$ due to the proximity of spin and charge instabilities \cite{PhysRevB_SDWCDW}, while at the overdoped region it is likely to pin to a commensurate vector of $q_{\rm CO}$ $\sim0.166~r.l.u.$. The Fermi-surface (FS) instabilities induced by the Coulomb interactions may lead to charge fluctuations in the sense of perturbation theory\,\cite{dalla2016friedel}. Accordingly, we have calculated the Lindhard function for LSCO and tracked the doping dependence of the FS nesting vector along the $(H, 0)$ direction. The nesting vector shows a non-monotonic behavior with a dip at $p \sim 0.3$ due to the Lifshitz transition from hole-like FS to electron-like FS (see \emph{Supplementary} Fig. S8). This cannot account for the nearly doping-independent wavevector at (0.166, 0), suggesting that the FS instability is an unlikely route to explain the CO in the overdoped regime. 

We then consider nonperturbatively strong correlations of electrons, which have been found to persist in the extremely overdoped cuprates \,\cite{paramagnon}. With strong correlations, charge fluctuations are frozen at the Mott insulating phase and become negligible when the valence band is close to empty. Therefore, a dome-like structure is naturally expected in the strongly correlated Hubbard model upon doping (see \emph{Supplementary} Fig. S10). However, the nature of the particle-hole excitations in the Hubbard model causes the leading instability at $(0.5,0.5)$, inconsistent with the observed charge mode. To explain the observed wavevector, we argue that it may be pinned by additional, longer-range interactions, which can be either remnant Coulomb repulsion or attraction mediated by bosonic modes, such as phonons \,\cite{chen2021anomalously}. In small-cluster simulations, we indeed observe the rise of a $q$ = (0.25, 0) [the smallest momentum accessible in the cluster] charge mode near 50\% doping (see \emph{Supplementary} Fig. S11). The dominance of this mode over the $(0.5,0.5)$ instability of the Hubbard model requires both relatively strong on-site and near-neighbor e-ph coupling besides the local Hubbard interaction. If the strong e-ph coupling is necessary for the experimentally observed CO, the absence of softening of phonons in our RIXS measurements suggests that other phonon mode(s) might be relevant, such as very low energy phonons observed in underdoped YBa$_2$Cu$_3$O$_{6.6}$ \cite{YBCOIXS} and La$_{2-x}$Ba$_x$CuO$_{4}$ \cite{LBCOIXS}, that should be looked for in the future.

We note that recent work has realized superconductivity in highly overdoped La$_{2-x}$Ca$_x$CuO$_4$ thin films with doping levels up to $x=0.5$ \cite{Kime2106170118}. This invites future experiments to reveal the relationship between SC and CO in the extremely overdoped regime. Moreover, our findings resemble the layered dichalcogenides Cu$_x$TiSe$_2$ \cite{NP2006} and infinite-layer nickelate \cite{NP2022}, where SC appears in the vicinity of charge order phase. This implies unusual competition between superconductivity and charge order states in broad correlated electron systems and motivates future investigations on unconventional superconductors.

\begin{acknowledgments}
We thank M. Grill, E. Huang, H. Yao, J. Zhang, X. J. Zhou, Y. Li, F. Wang, N. L. Wang, J. Feng, Z. Y. Weng for enlightening discussions. Y. Peng acknowledges the financial support from the National Natural Science Foundation of China (Grant No. 11974029) and the Ministry of Science and Technology of China (Grant No. 2019YFA0308401). Y. Xie acknowledges the financial support from the National Natural Science Foundation of China (Grant No. 12074334). I. E. and Y. W. acknowledge support from the National Science Foundation (NSF) award DMR-2038011. The RIXS experimental data were collected at beamline 41A of the National Synchrotron Radiation Research Center (NSRRC) in Hsinchu 30076, Taiwan. The REIXS experimental data were collected at the UE46-PGM1 beamline of Bessy-II (Helmholtz-Zentrum Berlin, Germany). 
\end{acknowledgments}

\end{document}